\documentclass[12pt]{article}
\usepackage{amssymb,amsmath,graphicx,color}
\begin{document}

\vspace{9mm}

\begin{center}
{{{\Large \bf Coupling between M2-branes and Form Fields}
}\\[17mm]
Yoonbai Kim,~~O-Kab Kwon,~~Hiroaki Nakajima,~~D.~D. 
Tolla$^{1}$\\[3mm]
{\it Department of Physics,~BK21 Physics Research Division,
~Institute of Basic Science,\\
$^{1}$University College\\
Sungkyunkwan University, Suwon 440-746, Korea}\\
{\tt yoonbai,~okab,~nakajima,~ddtolla@skku.edu} }
\end{center}

\vspace{20mm}

\begin{abstract}
In the context of low-energy effective theory of multiple M2-branes,
we construct the interaction terms between the world-volume fields of
M2-branes and the antisymmetric tensor fields of three- and six-forms.
By utilizing the compactification procedure, we show coincidence between
the dimensionally reduced coupling and the R-R
coupling to D-branes in type II string theory. We also discuss that
a cubic term proportional to six-form field reproduces the
quartic mass-deformation term in the world-volume theory of multiple
M2-branes.
\end{abstract}

\newpage


\setcounter{equation}{0}
\section{Introduction}\label{sec1}

To the leading order, the low-energy dynamics of a stack of $N$ parallel
D-branes is described by the super Yang-Mills action with
U($N$) gauge symmetry and the couplings to the bulk fields.
In string theory, D-branes carry R-R charges
and couple to R-R fields.
The form of interaction is given by
Wess-Zumino(WZ)-type action~\cite{Li:1995pq,Douglas:1995bn,Myers:1999ps}.
Unlike the case of a single D$p$-brane where
it couples only to R-R fields of rank $p+1$ or less, a stack of $N$
parallel D-branes couples to all even-form R-R fields in type IIB
string theory and to all odd-form R-R fields in type IIA string theory.

Analogous to the D-branes of string theory, in M-theory, we have the
M2- and M5-branes and the corresponding three- and six-form fields.
About their dynamics, the construction of the world-volume action of
multiple M-branes as well as their coupling to the form fields
is more difficult than that of multiple D-branes.
Recently the world-volume description of low-energy dynamics of multiple
M2-branes is available, which is Bagger-Lambert-Gustavsson (BLG) theory
with ${\cal N}=8$ supersymmetry and SU(2)$\times$SU(2) gauge
symmetry~\cite{Bagger:2006sk,Gustavsson:2007vu} and
Aharony-Bergman-Jafferis-Maldacena (ABJM) theory with ${\cal N}=6$
manifest supersymmetry and U($N$)$\times$U($N$) (or SU($N$)$\times$SU($N$)
gauge symmetry)~\cite{Aharony:2008ug}.
Once the action of world-volume fields is obtained, reproduction of its string
theory limit is an attractive research direction.
The reduction to type IIA string theory of the BLG theory upon circle
compactification in the direction transverse to the M2-brane has been
achieved in Ref.~\cite{Mukhi:2008ux,Ezhuthachan:2009sr}.

Despite an overwhelming progress in the understanding of the world-volume
action of multiple M2-branes, a little have been done to couple them to
the bulk form fields~\cite{Li:2008eza,Ganjali:2009kt}.
Therefore, it is intriguing to
construct the mutual interaction between the three- or six-form fields and
the world-volume fields in the context of BLG and ABJM theories.
It is the main objective of this paper to make a proposal for the action
which describes the interaction between the M2-branes and the form fields
of arbitrary transverse field dependence in the context of BLG theory
and to verify the proposal 
by reducing it to a similar interaction term in type IIA string theory
through circle compactification.

To see the importance of these interaction terms, we make a quick comparison
between M2-brane dynamics and the corresponding D2-brane in string theory.
In analogy with D2-brane dynamics, in the presence of nonvanishing
three-form and dual six-form fields, the low energy dynamics of multiple
parallel M2-branes is expected to be described by both their world-volume
action, the BLG action $S_{{\rm BLG}}$ in our case, and the coupling between
M2-branes and form fields $S_{C}$,
\begin{align}
S_{11}=S_{{\rm BLG}}+S_{C}.
\label{11d}
\end{align}
On the other hand, in the type IIA superstring theory the D2-brane action
possesses the Dirac-Born-Infeld (DBI) type (or Yang-Mills type in low energy
limit) world-volume action of $N$ parallel D2-branes $S_{{\rm DBI}}$ and
the R-R coupling\footnote{We put tildes for the fields and the parameters
in string theory.} $S_{\tilde C}$,
\begin{align}
S_{10}=S_{{\rm DBI}}+S_{\tilde C}.
\end{align}
In nontrivial background, the DBI action is given in terms of the
gauge-invariant field strength $\tilde F_{\mu\nu}+\tilde B_{\mu\nu}$,
where $\tilde F_{\mu\nu}$ is the field strength of the U($N$) gauge field
and $\tilde B_{\mu\nu}$ is the NS-NS two-form field. In the absence of
$S_C$, it have been verified that, upon circle compactification,
the action $S_{{\rm BLG}}$ reduces to Yang-Mills matter action composed of
only $\tilde F_{\mu\nu}$. In this paper we show that, 
after the compactification,
the presence of $S_C$ not only produces $S_{\tilde C}$ it also gives
the missing $\tilde B_{\mu\nu}$ piece of the DBI action in the NS-NS
background.

Some particular configuration of form fields coupled to D-branes or M-branes
can be regarded as the mass deformation of world-volume
theories~\cite{Polchinski:2000uf,Bena:2000zb}.
The (SUSY-preserving) mass deformation of BLG theory is explicitly constructed
in~\cite{Gomis:2008cv,Hosomichi:2008qk}, which contains the quartic
coupling among scalar fields as well as the quadratic mass terms.
We show that the WZ-type coupling with particular configuration of form fields
reproduces this quartic coupling.

The remaining part of this paper is organized as follows. In section 2 we put
forward our proposal for the three- and six-form couplings to multiple
M2-branes in BLG theory. Our proposal is made in parallel with the known
multiple D2-brane coupling to R-R forms. In section 3 we show that
the circle compactification of the action reproduces the corresponding action
in ten-dimensional IIA string theory. In section 4
we single out a particular term in the six-form coupling and show that,
with a proper choice of the constant background form field, it gives rise
to the quartic mass deformation of the BLG theory. Section 5 is
devoted to conclusions and discussions.

\setcounter{equation}{0}
\section{Coupling between M2-branes and Form
Fields in BLG Theory}\label{sec2}

In string theory the coupling of any D$p$-brane to R-R form
fields $\tilde{C}_{(n)}$ is given by the WZ-type action
as~\cite{Li:1995pq,Douglas:1995bn,Myers:1999ps}
\begin{align}
S_{{\tilde C}}=\mu_p\int_{p+1}{\rm STr}
\left(P\left[e^{i{\tilde\lambda} {\rm
i}_{\tilde X} {\rm i}_{\tilde X}}\sum {\tilde C}_{(n)}
e^{\tilde{B}}\right]e^{{\tilde\lambda} {\tilde F}}\right),
\label{MCSA1}
\end{align}
where $\mu_p$ is R-R charge of the D$p$-brane, ${\tilde\lambda}=2\pi l_{\rm
s}^2$ is the string scale, ${\tilde X}$ is the transverse scalar field,
$\tilde{B}$ is the NS-NS two-form field, and
${\tilde F}=d{\tilde A}$ is field strength of the gauge field ${\tilde A}$.
The trace is taken over the gauge indices, $P[...]$ denotes pullback,
the summation is taken over all the R-R forms, and ${\rm i_\Phi}$
represents an interior product by $\Phi^i$, explicitly written
\begin{align}
{\rm
i_\Phi}{\tilde C}^{(n)}=\frac{1}{(n-1)!}\,
\Phi^i {\tilde C}_{ii_1...i_{n-1}}dx^{i_1}\wedge\,...\,\wedge dx^{i_{n-1}}.
\end{align}
For later convenience we expand the exponential $e^{i{\tilde\lambda}
{\rm i_{\tilde X} i_{\tilde X}}}$ and write explicitly  the first few terms in
the case of $p=2$,
\begin{align}
S_{{\tilde C}}=\mu_2\int&
\frac{1}{3!} dx^\mu\wedge dx^\nu\wedge dx^\rho~
{\rm STr}\Big\{  {\tilde C}_{\mu\nu\rho}
+3\lambda {\tilde C}_{\mu\nu i}{\tilde D}_\rho {\tilde X}_i
+3\lambda^2 {\tilde C}_{\mu ij}{\tilde D}_\nu {\tilde X}_i{\tilde D}_\rho
{\tilde X}_j
\nonumber\\
& +\lambda^3 {\tilde C}_{ijk}{\tilde D}_\mu {\tilde X}_i {\tilde D}_\nu
{\tilde X}_j {\tilde D}_\rho {\tilde X}_k -3i\lambda^2 {\tilde C}_{\mu ij}
{\tilde X}_i{\tilde X}_j F_{\nu\rho} -3i\lambda^3{\tilde C}_{ijk}
{\tilde X}_i{\tilde X}_j {\tilde D}_\mu {\tilde X}_k F_{\nu\rho}
\nonumber \\
&-i\lambda{\tilde C}_{\mu\nu\rho ij}{\tilde X}_i{\tilde X}_j
-3i \lambda^2{\tilde C}_{\mu\nu ijk}{\tilde X}_i{\tilde X}_j
{\tilde D}_\rho {\tilde X}_k -3i\lambda^3{\tilde C}_{\mu ijkl}{\tilde X}_i
{\tilde X}_j {\tilde D}_\nu {\tilde X}_k {\tilde D}_\rho {\tilde X}_l
\nonumber \\
&-i\lambda^4 {\tilde C}_{ijklm}{\tilde X}_i {\tilde X}_j
{\tilde D}_\mu{\tilde X}_k
{\tilde D}_\nu {\tilde X}_l {\tilde D}_\rho {\tilde X}_m -\frac{3}{2}\lambda^3
{\tilde C}_{\mu ijkl}{\tilde X}_i{\tilde X}_j {\tilde X}_k{\tilde X}_l
F_{\nu\rho}
\nonumber \\
&-\frac{3}{2}\lambda^4{\tilde C}_{ijklm}{\tilde X}_i{\tilde X}_j
{\tilde X}_k{\tilde X}_l{\tilde D}_\mu{\tilde X}_m F_{\nu\rho}
+\cdots \Big\},\label{MCSA}
\end{align}
where ${\tilde X}_i\,\, (i=1,2,...,7)$ are seven transverse adjoint
scalar fields with ${\tilde D}_\mu {\tilde X}_i = \partial_\mu {\tilde X}_i
+ i [A_\mu,\, {\tilde X}_i]$. Here we omitted
the ${\tilde C}\wedge \tilde{B}$-terms
for simplicity. In M-theory we
naturally expect a similar coupling between the M-branes (M2- and
M5-branes) and the antisymmetric form fields (the three-form field
$C_{(3)}$ and the dual six-form field
$C_{(6)}$)~\cite{Townsend:1996xj}. Since $\tilde B$ is already a part of
$C_{(3)}$, $C_{8ij}\sim \tilde{B}_{ij}$ for the compactified eighth
direction, the expected action involves the interaction between the
form fields and the world-volume fields. In the following,
we will consider the BLG theory and construct an analogue
of the coupling between the world-volume fields and the form fields.
Instead of the original formulation based on
three-algebra, we employ a familiar gauge theory
formulation~\cite{VanRaamsdonk:2008ft,Benna:2008zy}.

We begin with the BLG theory with eight transverse bi-fundamental
scalar fields
$X_{I}$ $(I=1,2,...,8)$ and two gauge fields $A$ and ${\hat A}$ of
SU(2)$\times$SU(2) gauge symmetry.
The bosonic part of the action is
\begin{align}
S_{\rm bos}=S_X+S_{{\rm CS}}+S_{C}.
\end{align}
The first two are well established and are given by
\begin{align}
S_X&=\int d^3x\, {\rm Tr}\Big[-(D_\mu X_I)^\dagger
D^\mu X_I-\frac{32\pi^2}{3k^2}X_{IJK}X_{IJK}^\dagger\Big],
\label{Xac} \\
S_{{\rm CS}}&=\frac k{4\pi}\int d^3x\, \epsilon^{\mu\nu\rho}{\rm Tr}
\Big(A_\mu\partial_\nu A_\rho+\frac{2i}3A_\mu A_\nu A_\rho
-\hat A_\mu\partial_\nu \hat A_\rho-\frac{2i}3\hat A_\mu \hat A_\nu
\hat A_\rho\Big),
\label{CSac}
\end{align}
where $k$ is the Chern-Simons level and we have used the notation
 \begin{align}
D_{\mu}X_I=\partial_\mu X_I+iA_\mu-iX_I \hat
A_\mu, \qquad X_{IJK}\equiv X_{[I}X^\dagger_J X_{K]}.
\end{align}

Due to T-duality between type IIA and IIB string theories,
one can restrict possible
interaction terms between D-branes and R-R form fields.
The exponential factor of the gauge field ${\tilde F}$ in the R-R coupling
\eqref{MCSA1} is introduced along the line of open string tadpole
computation and is compatible with the
T-duality~\cite{Li:1995pq,Douglas:1995bn,Bergshoeff:1996cy,Myers:1999ps}.
In addition gauge invariance on the D-brane requires
$\tilde{B}+{\tilde \lambda}{\tilde F}$
combination and it justifies the exponential factor of the
NS-NS two-form field in \eqref{MCSA1}.
Unlike the superstring theories, however, there seems no concrete guideline
for the interactions between M-branes and form fields in M-theory yet.
The candidate for the WZ-type coupling between M2-branes and form
fields, which is linear in the form fields, is
\begin{align}
S_{{\tilde C}} &=\int_{2+1}{\rm Tr}\Big(\mu_2 P [C_{(3)}]+\mu'_2 P[
\langle{\rm i}_X{\rm i}_X{\rm i}_X\rangle C_{(6)}]\Big)
\label{liC}\\
&=\int\frac{1}{3!}\, d^3x\,\epsilon^{\mu\nu\rho}~{\rm Tr}\bigg\{
\frac{\mu_2}{2} \Big[\frac12 \big(\hat C_{\mu\nu\rho}+C_{\mu\nu\rho}\big)
+3\lambda C_{\mu\nu I}
(D_\rho X_I)^\dagger \nonumber\\
& \hspace{43mm}+\frac32\lambda^2 \Big(\hat C_{\mu IJ}(D_\nu X_I)^\dagger
D_\rho X_J+C_{\mu IJ}D_\nu X_I(D_\rho X_J)^\dagger\Big)
\nonumber\\
& \hspace{43mm} +\lambda^3 C_{IJK}(D_\mu
X_I)^\dagger D_\nu X_J (D_\rho X_K)^\dagger \Big] +({\rm c.c.})
\nonumber\\
& \hspace{33mm} + \frac{\mu'_2}{2} \Big[C_{\mu\nu\rho
IJK}X_{IJK}^\dagger  +\frac {3}2\lambda
\hat C_{\mu\nu IJKL}\langle\hspace{-0.7mm}\langle X_{IJK}^\dagger
D_\rho X_L\rangle\hspace{-0.7mm}\rangle\nonumber\\
 & \hspace{43mm} +\frac {3}2\lambda C_{\mu\nu IJKL}
\langle\hspace{-0.7mm}\langle X_{IJK} (D_\rho X_L)^\dagger\rangle
\hspace{-0.7mm}\rangle\nonumber\\
\label{MCSX}
& \hspace{43mm} +3\lambda^2 C_{\mu
IJKLM}\langle\hspace{-0.7mm}\langle
X_{IJK}^\dagger D_\nu X_L (D_\rho X_M)^\dagger
\rangle\hspace{-0.7mm}\rangle
\nonumber\\
& \hspace{43mm} +\frac12\lambda^3 \hat C_{IJKLMN}
\langle\hspace{-0.7mm}\langle X_{IJK}^\dagger
 D_\mu X_L (D_\nu X_M)^\dagger D_\rho X_N
\rangle\hspace{-0.7mm}\rangle \nonumber\\
& \hspace{43mm}+\frac12\lambda^3 C_{IJKLMN}
\langle\hspace{-0.7mm}\langle X_{IJK}
 (D_\mu X_L)^\dagger D_\nu X_M (D_\rho X_N)^\dagger
\rangle\hspace{-0.7mm}\rangle \Big]+({\rm c.c.}) \bigg\},
\end{align}
where $\mu_{2}$ is M2-brane tension, $\lambda=2\pi  l_{{\rm P}}^{3/2}$
with Planck length $l_{{\rm P}}$,
and $\mu'_{2}= \beta \lambda\mu_{2}$. 
Dimensionless parameter $\beta$ will be fixed by requiring that
when reduced to ten dimensions this action reproduces the correct
D2-brane coupling to R-R and NS-NS form fields in type IIA superstring
theory. 
$\langle{\rm i}_X{\rm i}_X{\rm i}_X\rangle$ denotes interior products by
$X_{IJK}$ and its Hermitian conjugate $X^\dagger_{IJK}$ in gauge invariant
manner.
We introduce the notation
$\langle\hspace{-0.7mm}\langle\,...\rangle\hspace{-0.7mm}\rangle$ to
symmetrize objects inside the trace, for instance,
\begin{align}
&\langle\hspace{-0.7mm}\langle X_{IJK}^\dagger D_\nu X_L (D_\rho X)^\dagger_M
\rangle\hspace{-0.7mm}\rangle
\nonumber \\
&\hspace{10mm}
=\frac{1}{3}\left[ X_{IJK}^\dagger D_\nu X_L (D_\rho X_M)^\dagger +
(D_\nu X_L)^\dagger X_{IJK} (D_\rho X_M)^\dagger +
(D_\nu X_L)^\dagger D_\rho X_M X_{IJK}^\dagger\right].
\nonumber
\end{align}
We also note that the different powers of the Planck length $l_{\rm
P}$ in front of some of the terms in  the action are chosen based on
dimension counting and the $(2\pi )^n$ factors are inserted to
mimic the similar factors in ten dimensions.

Now we recall that the
scalar fields $X_I$ transform in the bi-fundamental representation of
the gauge group SU(2)$\times$SU(2), while $X_I^\dagger$ transform in
the anti-bi-fundamental representation. In order to have gauge invariance,
we  realize that the $C_{\mu\nu\rho IJK}$ and all the
other antisymmetric tensor fields with odd number of transverse
indices should be in the bi-fundamental representation.
For the same reason $ C_{\mu\nu\rho}$ and
all the other antisymmetric tensor fields $ C$ with even number of the
transverse indices should be in the adjoint of the left ${\rm {SU}(2)}$,
while $\hat C_{\mu\nu\rho}$ and
all the other antisymmetric tensor fields $\hat C$ with even number of the
transverse indices should be in the adjoint of the right SU(2). 
With these transformation rules all the terms in (\ref{MCSX}) 
are gauge invariant.

In Ref.~\cite{Li:2008eza}, the authors proposed the
WZ-type couplings in M-theory in terms of 3-algebras with Euclidean and
Lorentzian metrics. Some of terms of their proposal for the
WZ-type action resemble those in our action (\ref{MCSA}), however,
in Ref.~\cite{Li:2008eza} the authors assumed that the three- and six-form
fields do not depend on the transverse scalar fields and
so transform trivially under the gauge transformation. To get a gauge
invariant action they introduced symmetrized constant tensors originated from
symmetrized trace of generators with 3-algebra indices.
For a specific representation of the 3-algebra, the symmetrized tensors
satisfying the gauge invariance of the action were obtained
as functions of structure constant of the 3-algebra.
After that, ten-dimensional WZ-type action was obtained by using
the Higgs mechanism proposed in Ref.~\cite{Mukhi:2008ux}.
Since the resulting action is composed of constant form fields and
symmetric tensors depending on specific representation of 3-algebra,
it is not clear to relate the results to the known WZ-type action
(\ref{MCSA1}) expressed by U(2)-adjoints.

\setcounter{equation}{0}
\section{Reduction from M-theory to IIA String Theory}\label{sec3}

In the previous section we constructed the analogue
of WZ-type coupling \eqref{MCSX} between M2-branes and form fields.
In this section we shall test and justify the obtained candidate
by comparing it with the R-R coupling  in string theory \eqref{MCSA}
by reducing it to the ten-dimensional type IIA superstring theory.
Specifically we expand the action \eqref{MCSA1} and compare
the obtained result (\ref{MCSA})
with the dimensionally reduced WZ-type action of M-theory \eqref{MCSX}.

According to the compactification procedure of Ref.~\cite{Mukhi:2008ux},
we can split the transverse scalars into trace and traceless parts,
\begin{align}
X_i &= {\check x}_i
+ i {\bf x}_i, \quad (i=1,2,...,7),
\nonumber \\
X_8&=\frac{v}{2}{\bf 1}+{\check x}_8
+ i {\bf x}_8,
\end{align}
where $\check x_I=x_I^4\frac{\bf 1}2$, ${\bf x}_I = x^\alpha_I
\frac{\sigma^\alpha}{2}$ $(\alpha=1,2,3)$, and
$v$ is a very large vacuum expectation value of ${\rm Tr} X_8$.
We also introduce
\begin{align}
A_\mu^{\pm}=\frac{1}{2}(A_\mu\pm\hat A_\mu),
\end{align}
and then $A_\mu^-$ becomes an
auxiliary gauge field which we can integrate out using its equation of
motion. We can rewrite the covariant derivatives as
\begin{align}\label{tcov}
D_\mu X_I=\tilde D_\mu X_I+i\{A^-_\mu,X_I\} ~~~{\rm with}~~~
\tilde D_\mu X_I=\partial_\mu X_I+i[ A^+_\mu,X_I].
\end{align}
In order to consider the type IIA limit for the coupling
proposed in (\ref{MCSX}), we have to obtain the form of covariant
derivatives for the transverse scalars $X_I$ in the limit.
Taking into account the contributions from  Chern-Simons term (\ref{CSac}),
the kinetic term for the transverse
scalar fields and the WZ-type terms, we will solve the equation
of motion for $A_\mu^-$
in the limit of a large vacuum expectation value $v$ and large
Chern-Simons level $k$,  with a fixed
$v/k$. It turns out that the leading term in the solution to $A^-_\mu$
is linear in $1/v$ and we can neglect every term containing $A^-_\mu$
unless it is multiplied by $v$ or $k$.
Keeping this in mind we rewrite the covariant derivatives for $X_8$
and the other transverse scalars $X_i$ as follows
\begin{align}\label{dmux8}
&D_\mu X_8=\partial_\mu {\check x}_8 +
iv(A_\mu^{-} + \frac{1}{v} \tilde D_\mu {\bf x}_8),\nonumber\\
&D_\mu X_i=\tilde D_\mu X_i.
\end{align}
Here we notice that the appearance of $A_\mu^-$ in the WZ-type action
is only through $D_\mu X_8$, which means it always appears in the combination
$(A_\mu^{-} + \frac{1}{v} \tilde D_\mu {\bf x}_8)$. 
$A_\mu^-$ in the $S_X+S_{{\rm CS}}$ also appears only in this combination 
as verified by Ref.~\cite{Mukhi:2008ux}. 
Therefore, applying the Higgs mechanism, we make
a shift of the gauge field
$A^-_\mu\to A^{-}_\mu - \frac{1}{v} D^{+}_\mu x_8$ to eliminate
the traceless part of the eighth scalar field ${\bf x}_8$ in the resulting
Lagrangian. With this shift the covariant derivative of $X_8$ becomes
\begin{align}\label{dmux8s}
D_\mu X_8=\partial_\mu {\check x}_8 +
iv A_\mu^{-} .
\end{align}
We also adopt the Higgs rule in Ref.~\cite{Ezhuthachan:2009sr}
for the covariant derivatives of the scalars $X_i$ and rewrite
them as
\begin{align}
D_\mu X_i\to i\tilde D_\mu\tilde X_i,~~~~~~~~~~ (D_\mu X_i)^\dagger\to
-i\tilde D_\mu\tilde X_i,
\end{align}
where $\tilde X_i=\check x_i+{\bf x}_i$ are U(2) adjoint scalars.

Now the action for the scalar fields (\ref{Xac}) 
will take the following simple form
\begin{align}
S_X=\int d^3x \, {\rm Tr}\Big(-\tilde D_\mu \tilde X_i \tilde D^\mu
\tilde X_i-\partial_\mu \check x_8\partial^\mu\check x_8
-v^2A_\mu^-A^{-\mu}-V_{\rm bos}\Big)+{\cal O}\big(\frac 1v\big),
\end{align}
where $V_{\rm bos}$ is the potential term.
This term will not be affected by the WZ-type action. Therefore,
we will not write it explicitly except in the final result.
The Chern-Simons action (\ref{CSac}) also reduces to
\begin{align}
S_{{\rm CS}}=\frac k{2\pi}\int d^3x\, \epsilon^{\mu\nu\rho}{\rm Tr}
\Big(A^-_\mu\tilde F_{\nu\rho}\Big)+{\cal O}\big(\frac 1v\big)~~~~~~
{\rm with}~~~{\tilde F}_{\mu\nu}=\partial_\mu A^+_\nu-\partial_\nu A^+_\mu
+i[A^+_\mu ,A^+_\nu].
\end{align}
For our immediate purpose of solving the $A_{\mu}^-$ equation of motion,
we write only the part of the WZ-type action that involves $A_{\mu}^-$ 
explicitly, leaving the remaining part implicit until the end of this section 
\begin{align}
S_{C} =&\int\frac{1}{3!}\,d^3x\,\epsilon^{\mu\nu\rho}~{\rm Tr}\Bigg\{
\frac{\mu_2}{2} \Big[\frac12\big(\hat C_{\mu\nu\rho}+C_{\mu\nu\rho}\big)
-3i\lambda C_{\mu\nu i}
\tilde D_\rho \tilde X_i+3\lambda C_{\mu\nu 8}(\partial_\rho {\check x}_8 -
iv A_\rho^{-} )\nonumber\\
&+\frac32\lambda^2 \big(\hat C_{\mu ij}+C_{\mu ij}\big)\tilde D_\mu
\tilde X_i \tilde D_\nu
\tilde X_j -\frac32i\lambda^2 \hat C_{\mu i8}\Big(\tilde D_\nu \tilde X_i
(\partial_\rho {\check x}_8 +iv A_\rho^{-} )
-(\partial_\rho {\check x}_8 -iv A_\rho^{-} ) \tilde D_\nu \tilde X_i\Big)
\nonumber\\
&+\frac32i\lambda^2 C_{\mu i8}\Big(\tilde D_\nu \tilde X_i (\partial_\rho
{\check x}_8 -iv A_\rho^{-} )
-(\partial_\rho {\check x}_8 +iv A_\rho^{-} ) \tilde D_\nu \tilde X_i\Big)
-i\lambda^3 C_{ijk}\tilde D_\mu \tilde X_i \tilde D_\nu \tilde X_j
\tilde D_\rho \tilde X_k\nonumber\\
& +\lambda^3 C_{ij8}\Big(\tilde D_\mu \tilde X_i
\tilde D_\nu \tilde X_j (\partial_\rho {\check x}_8 -
iv A_\rho^{-} )-\tilde D_\mu \tilde X_i (\partial_\rho {\check x}_8 +
iv A_\rho^{-} ) \tilde D_\nu \tilde X_j\nonumber\\
&+(\partial_\rho {\check x}_8 -
iv A_\rho^{-} ) \tilde D_\mu \tilde X_i \tilde D_\nu \tilde X_j\Big)\Big]
+({\rm c.c.})
\nonumber\\
& + \frac{\mu'_2}{2} \Big[C_{\mu\nu\rho
ij8}X_{ij8}^\dagger+\cdots  +\frac {3}2i\lambda\Big(
\hat C_{\mu\nu ij8k}\langle\hspace{-0.7mm}\langle X_{ij8}^\dagger \tilde
D_\rho \tilde X_k\rangle\hspace{-0.7mm}\rangle
-C_{\mu\nu ij8k}\langle\hspace{-0.7mm}\langle X_{ij8} \tilde D_\rho
\tilde X_k\rangle\hspace{-0.7mm}\rangle\Big)+\cdots\nonumber\\
&+3\lambda^2 C_{\mu
ij8kl}\langle\hspace{-0.7mm}\langle
X_{ij8}^\dagger \tilde D_\nu \tilde X_k \tilde D_\rho \tilde X_l
\rangle\hspace{-0.7mm}\rangle+\cdots
  +\frac12 i\lambda^3 \hat C_{ij8klm}
\langle\hspace{-0.7mm}\langle X_{ij8}^\dagger
 \tilde D_\mu \tilde X_k \tilde D_\nu \tilde X_l \tilde D_\rho \tilde X_m
\rangle\hspace{-0.7mm}\rangle+\cdots\nonumber\\
&  +\frac12 i\lambda^3 C_{ij8klm}
\langle\hspace{-0.7mm}\langle X_{ij8}
 \tilde D_\mu \tilde X_k \tilde D_\nu \tilde X_l \tilde D_\rho \tilde X_m
\rangle\hspace{-0.7mm}\rangle+\cdots\Big] +({\rm c.c.}) \Bigg\}+{\cal O}
\big(\frac 1v\big).
\label{MCSX2}
\end{align}
For the terms proportional to $\mu_2'$ in \eqref{MCSX2}, we kept
the leading terms proportional to $v$
but neglected all the higher order terms, ${\cal O}(k/v,k/v^2,...)$. The
reason is that, as we pointed out before,  $\mu_2'\sim \beta$ and
we will show shortly that the numerical factor $\beta$ is of the order
of $1/k$ which is of order of $1/v$.

The variation of the action with respect to $A^-_\mu$ gives 
\begin{align}
0=&{\rm Tr}\bigg\{\bigg[-2v^2A^{-\mu}+\frac k{2\pi}
\epsilon^{\mu\nu\rho}
{F}_{\nu\rho}+{\mu_2 v\lambda}\epsilon^{\mu\nu\rho}
\Big[-\frac i4(C_{\nu\rho8}-C^\dagger_{\nu\rho8})\nonumber\\
&~~~+\frac\lambda 4
\Big((\hat C_{\nu i8}+C_{\nu i8})\tilde D_\rho \tilde X_i
+\tilde D_\rho \tilde X_i(\hat C_{\nu i8}+C_{\nu i8})\Big)
-\frac {i\lambda^2}{12}\Big((C_{ij8}-C_{ij8}^\dagger)
\tilde D_\nu \tilde X_i\tilde D_\rho \tilde X_j
\nonumber \\
&~~~+\tilde D_\nu
\tilde X_i(C_{ij8}-C_{ij8}^\dagger)\tilde D_\rho \tilde X_j
+\tilde D_\nu
\tilde X_i\tilde D_\rho \tilde X_j(C_{ij8}-C_{ij8}^\dagger)\Big)\Big]\bigg]
\delta A_\mu^-\bigg\}.
\label{EOM}
\end{align}
After integrating out $A_\mu^-$, we recall that the two SU(2) groups will be
identified and as a result $C$ and $\hat C$ become the same. In addition,
if we identify the NS-NS two-form field in type II string theory as
\begin{align}\label{Bmunu}
\tilde B_{P\!R}=\frac 14\Big[C_{P\!R 8}+C^\dagger_{P\!R 8}-i(C_{P\!R 8}
-C^\dagger_{P\!R 8})\Big],
\end{align}
then the quantity inside square bracket in (\ref{EOM}) gives
the traceless part of pullback of $\tilde B_{\mu\nu}$,
\begin{align}\label{PBmunu}
P[\tilde B_{\mu\nu}]=\tilde B_{\mu\nu}+\lambda \tilde B_{\mu i}
\tilde D_\nu \tilde X_i+\lambda \tilde D_\nu \tilde X_i\tilde B_{\mu i}
+\frac{\lambda^2}3\Big(\tilde B_{ij}
\tilde D_\nu \tilde X_i\tilde D_\rho \tilde X_j+\tilde D_\nu \tilde X_i
\tilde B_{ij}\tilde D_\rho \tilde X_j+\tilde D_\nu \tilde X_i\tilde D_\rho
\tilde X_j\tilde B_{ij}\Big).
\end{align}
On the other hand, since $\delta A_\mu^-$ is traceless, the product of
the trace part of $P[\tilde B_{\mu\nu}]$ and  $\delta A_\mu^-$ is also
traceless. Therefore, the $A_\mu^-$ equation of motion 
is simplified as
\begin{align}
0={\rm Tr}\bigg\{\Big[-2v^2A^{-}_\mu +\frac k{2\pi}\epsilon^{\mu\nu\rho}\Big(
{F}_{\nu\rho}+{\mu_2 v\lambda}\frac{2\pi}k
P[\tilde B_{\nu\rho}]\Big)\Big]\delta A_\mu^-\bigg\}.
\label{EOM2}
\end{align}
Noting that the Yang-Mills coupling constant $g_{{\rm YM}}$
in the effective action of two D2-branes, 
the dimensionless string coupling constant $g_{{\rm s}}$, and 
the string scale $\tilde\lambda$ are given by 
\begin{align}
g_{{\rm YM}}=\frac {2\pi v}k,\qquad 
g_{{\rm s}}=g^2_{{\rm YM}} l_{{\rm s}}, 
\qquad \tilde\lambda=2\pi l_{{\rm s}}^2 \label{list}
\end{align}
with $\mu_{2}\lambda=
\frac 1{4\pi^2 l_{\rm P}^3}{2\pi l_{\rm P}^{3/2}}$ and 
$l_{\rm P}=g_{{\rm s}}^{1/3}l_{{\rm s}}$, $A^-_\mu$ in (\ref{EOM2}) is 
\begin{align}\label{Amu}
A^{-}_\mu = \frac{1}{2g_{{\rm YM}} \tilde \lambda v}\,\epsilon_\mu^{~\nu\rho}
\Big(P[B_{\nu\rho}]+\tilde\lambda{ F}_{\nu\rho}\Big),
\end{align}
where  $B_{\mu\nu}$ is
the traceless part of $\tilde B_{\mu\nu}$.

The gauge singlet scalar ${\check x}_8$ is 
dualized to a U(1) gauge field by replacing
\begin{align}\label{tilx8}
\partial_\mu{\check x}_8 = \frac{1}{2 g_{{\rm YM}}}\,
\epsilon_{\mu\nu\rho}{\check F}^{\nu\rho},
\end{align}
where ${\check F}_{\nu\lambda}$ is a U(1) gauge field strength.
The action  for the scalar fields $S_X$ (\ref{Xac}) is now given by
\begin{align}
S_X=\int d^3x\, {\rm Tr}\left[-\tilde D_\mu \tilde X_i \tilde D^\mu
\tilde X_i-\frac{1}{2 g^2_{{\rm YM}}}\check F_{\mu\nu}\check F^{\mu\nu}
+\frac{1}{2 g^2_{{\rm YM}}}\Big(F_{\mu\nu}+\frac 1{\tilde\lambda}
P[B_{\mu\nu}]\Big)^2-V_{\rm bos}\right]+{\cal O}\big(\frac 1v\big),
\label{SX10}
\end{align}
while the Chern-Simons action (\ref{CSac}) is given by
\begin{align}
S_{{\rm CS}}=\int d^3x\, {\rm Tr}\left[-\frac 1{g^2_{{\rm YM}}}\big(F_{\mu\nu}
+\frac 1{\tilde\lambda} P[B_{\mu\nu}]\big) F^{\mu\nu}\right]
+{\cal O}\big(\frac 1v\big)
\label{SCS10}.
\end{align}

By introducing the U(2) adjoint antisymmetric three-form fields
\begin{align}
\tilde C_{PRS}=\frac 12\Big[C_{PRS}+C_{PRS}^\dagger-i(C_{PRS}-C_{PRS}^\dagger)
\Big],
\end{align}
we have the $C_{(3)}$ part of the WZ-type action $S_C$, 
\begin{align}
S^{(3)}_{C} =\int\,d^3x\,\epsilon^{\mu\nu\rho}~{\rm Tr}\bigg\{
\frac{\mu_2}{3!} \Big[&\tilde C_{\mu\nu\rho} +3\lambda \tilde C_{\mu\nu i}
\tilde D_\rho \tilde X_i+3\lambda^2 \tilde C_{\mu ij}\tilde D_\mu \tilde X_i
\tilde D_\nu \tilde X_j+\lambda^3 \tilde C_{ijk}\tilde D_\mu \tilde X_i
\tilde D_\nu \tilde X_j\tilde D_\rho \tilde X_k\Big]
\nonumber\\
&+\mu_2\lambda\Big(vP[B_{\mu\nu}]A^-_\rho+ P[\check B_{\mu\nu}]\partial_\rho
\check x_8\Big)\bigg\}+{\cal O}\big(\frac 1v\big),
\label{SM3}
\end{align}
where $P[\check B_{\mu\nu}]$ is the trace part $P[\tilde B_{\mu\nu}]$.
Substituting (\ref{Amu})-(\ref{tilx8}) into (\ref{SM3}) 
for $A_\mu^-$ and $\partial_\rho \check x_8$ 
and taking into account the constants given in (\ref{list}), we obtain
\begin{align}
 S^{(3)}_{C} =\int d^3x~{\rm Tr}\bigg\{
\frac{\mu_2}{3!} \epsilon^{\mu\nu\rho}\Big[&\tilde C_{\mu\nu\rho}
+3\lambda \tilde C_{\mu\nu i}
\tilde D_\rho \tilde X_i+3\lambda^2 \tilde C_{\mu ij}\tilde D_\mu \tilde X_i
\tilde D_\nu \tilde X_j+\lambda^3 \tilde C_{ijk}\tilde D_\mu \tilde X_i
\tilde D_\nu \tilde X_j\tilde D_\rho \tilde X_k\Big]
\nonumber\\
&-\frac 1{g^2_{{\rm YM}}\tilde\lambda}\Big(\lambda P[B_{\mu\nu}]\big(F^{\mu\nu}
+\frac 1{\tilde\lambda}P[B^{\mu\nu}]\big)+\lambda P[\check B_{\mu\nu}]
\check F^{\mu\nu}\Big)\bigg\}+{\cal O}\big(\frac 1v\big)
\label{SM310}.
\end{align}

In order to match the mass
dimension of the ten-dimensional transverse scalar fields, we rescale
the scalar fields as $\tilde X_i\to \frac {\tilde X_i}{g_{\rm YM}}$.
Applying this rescaling to (\ref{SX10}), (\ref{SCS10}), (\ref{SM310})
and summing them, we get
\begin{align}
&S_{X} +S_{{\rm CS}} + S^{(3)}_{C} =
\nonumber \\
&~~~~~~~~~\int d^3x~{\rm Tr}\bigg\{\frac 1{g^2_{Y\!M}}
\Big[-\tilde D_\mu \tilde X_i \tilde D^\mu \tilde X_i-\frac 12
\Big(\check F_{\mu\nu}\check F^{\mu\nu}+\frac 2{\tilde\lambda}
\check F_{\mu\nu}P[\check B^{\mu\nu}]\Big)\nonumber\\
&~~~~~~~~~-\frac 12\Big(F_{\mu\nu}+\frac 1{\tilde\lambda}
P[B_{\mu\nu}]\Big)\Big(F^{\mu\nu}+\frac 1{\tilde\lambda}
P[B^{\mu\nu}]\Big)-\frac 12[\tilde X_i,\tilde X_j]
[\tilde X_i,\tilde X_j]\Big]\nonumber\\
&~~~~~~~~~+\frac{\mu_2}{3!} \epsilon^{\mu\nu\rho}\Big[\tilde C_{\mu\nu\rho}
+3\tilde\lambda \tilde C_{\mu\nu i}
\tilde D_\rho \tilde X_i+3\tilde\lambda^2 \tilde C_{\mu ij}
\tilde D_\mu \tilde X_i \tilde D_\nu \tilde X_j+\tilde\lambda^3
\tilde C_{ijk}\tilde D_\mu \tilde X_i \tilde D_\nu \tilde X_j
\tilde D_\rho \tilde X_k\Big]
\bigg\}\label{XCSM3},
\end{align}
where we used the explicit form of the scalar potential
$V_{\rm bos}$ in Ref.~\cite{Mukhi:2008ux}.

Next we turn to the $C_{(6)}$ part of $S_C$. After the aforementioned
rescaling $\tilde X_i\to \frac {\tilde X_i}{g_{\rm YM}}$, the Higgs rule
 for $X_{ij8}$~\cite{Ezhuthachan:2009sr} becomes
\begin{align}
X_{ij8}\to \frac v{4g^2_{{\rm YM}}}[\tilde X_i,\tilde X_j], ~~~~~~~~~
X^\dagger_{ij8}\to- \frac v{4g^2_{{\rm YM}}}[\tilde X_i,\tilde X_j]
\label{Xij8}.
\end{align}
We also introduce the U(2) adjoint antisymmetric six-form fields as
\begin{align}
\tilde C_{PQRSTV}=\frac 12\Big[C_{PQRSTV}+C_{PQRSTV}^\dagger
-i(C_{PQRSTV}-C_{PQRSTV}^\dagger)\Big]
\end{align}
and identify the antisymmetric five-form fields in type
IIA string theory as
\begin{align}
\tilde C_{PQRST}=\tilde C_{PQRST8}.
\end{align}

The first term in the $C_{(6)}$ Lagrangian in (\ref{MCSX2}) is given by
\begin{align}
\frac {\mu'_2}{2}{\rm Tr}\Big(C_{\mu\nu\rho IJK}X^\dagger_{IJK}
+C^\dagger_{\mu\nu\rho IJK}X_{IJK}\Big)
=-i\frac{3v\mu_2'}{8g^2_{{\rm YM}}}{\rm Tr}
\Big\{-i(C_{\mu\nu\rho ij8}-C^\dagger_{\mu\nu\rho ij8})
[\tilde X_i,\tilde X_j]\Big\}+{\cal O}\big(\frac 1v\big)\label{cij8}.
\end{align}
Since $[\tilde X_i,\tilde X_j]$ is traceless, the product
$(C_{\mu\nu\rho ij8}+C^\dagger_{\mu\nu\rho ij8})[\tilde X_i,\tilde X_j]$
is traceless. Therefore, we can freely include this term in the
last equation (\ref{cij8}) to get
\begin{align}
&\frac {\mu'_2}{2}{\rm Tr}\Big(C_{\mu\nu\rho IJK}X^\dagger_{IJK}
+C^\dagger_{\mu\nu\rho IJK}X_{IJK}\Big)
\nonumber \\
&~~~~~~~~~~~~~~=-i\frac{3v\mu_2'}{8g^2_{{\rm YM}}}{\rm Tr}
\Big(\tilde C_{\mu\nu\rho ij8}[\tilde X_i,\tilde X_j]\Big)
=-i\frac{\mu_2\tilde\lambda}{2}{\rm Tr}\Big(\tilde C_{\mu\nu\rho ij}
[\tilde X_i,\tilde X_j]\Big)\label{CIJK}.
\end{align}
In the second equality we have used 
$\mu_2'=\beta \lambda \mu_2$ and have chosen $\beta=\frac{4\pi}{3k}$
in order to match the coefficient with the coefficient of the corresponding
term in type IIA string theory in (\ref{MCSA}).
Following the same procedure, we can calculate the remaining terms
in the $C_{(6)}$ Lagrangian in (\ref{MCSX2}),
\begin{align}
&\frac {3}{2}\mu_2'\lambda{\rm Tr}\Big(C_{\mu\nu LIJK}
\langle\hspace{-0.7mm}\langle X_{IJK}^\dagger D_\rho X_L
\rangle\hspace{-0.7mm}\rangle\Big)+ {\rm c.c.}=-\frac {3i}{2}\mu_2
\tilde\lambda^2{\rm Tr}\Big(\tilde C_{\mu\nu ijk}\langle\hspace{-0.7mm}
\langle [\tilde X_i,\tilde X_j] \tilde D_\rho \tilde X_k
\rangle\hspace{-0.7mm}\rangle\Big),
\label{CIJKL}\\
&\frac {3}{2}\mu_2'\lambda^2{\rm Tr}\Big(C_{\mu LMIJK}\langle
\hspace{-0.7mm}\langle X_{IJK}^\dagger D_\nu X_L(D_\rho X_M)^\dagger
\rangle\hspace{-0.7mm}\rangle\Big)+{\rm c.c.}
\nonumber\\
&~~~~~~~~~~~~~~~~~~~~~~~~~~~~~=-\frac {3i}{2}\mu_2\tilde
\lambda^3{\rm Tr}\Big(\tilde C_{\mu ijkl}\langle\hspace{-0.7mm}\langle
[\tilde X_i,\tilde X_j] \tilde D_\nu \tilde X_k\tilde D_\rho \tilde X_l
\rangle\hspace{-0.7mm}\rangle\Big),
\label{CIJKLM}\\
&\frac {1}{2}\mu_2'\lambda^3{\rm Tr}\Big(C_{LMNIJK}\langle\hspace{-0.7mm}
\langle X_{IJK}^\dagger D_\mu X_L(D_\nu X_M)^\dagger D_\rho X_N\rangle
\hspace{-0.7mm}\rangle\Big)+{\rm c.c.}\nonumber\\
&~~~~~~~~~~~~~~~~~~~~~~~~~~~~~=-\frac {i}{2}\mu_2\tilde\lambda^4{\rm Tr}
\Big(\tilde C_{ijklm}\langle\hspace{-0.7mm}\langle [\tilde X_i,\tilde X_j]
\tilde D_\mu \tilde X_k \tilde D_\nu \tilde X_l\tilde D_\rho \tilde X_m
\rangle\hspace{-0.7mm}\rangle\Big).
\label{CIJKLMN}
\end{align}

Summing (\ref{XCSM3}) and (\ref{CIJK})-(\ref{CIJKLMN}),
we finally reach
\begin{align}
S_{{\rm tot}} =&\int d^3x~{\rm Tr}\bigg\{\frac 1{g^2_{Y\!M}}
\Big[-\tilde D_\mu \tilde X_i \tilde D^\mu \tilde X_i-\frac 12
\Big(\check F_{\mu\nu}\check F^{\mu\nu}+\frac 2{\tilde\lambda}
\check F_{\mu\nu}P[\check B^{\mu\nu}]\Big)\nonumber\\
&~~~~~~~~~~~~-\frac 12\Big(F_{\mu\nu}+\frac 1{\tilde\lambda}
P[B_{\mu\nu}]\Big)\Big(F^{\mu\nu}+\frac 1{\tilde\lambda}
P[B^{\mu\nu}]\Big)-\frac 12[\tilde X_i,\tilde X_j]
[\tilde X_i,\tilde X_j]\Big]\nonumber\\
&+\frac{\mu_2}{3!} \epsilon^{\mu\nu\rho}\Big[\tilde C_{\mu\nu\rho}
+3\tilde\lambda \tilde C_{\mu\nu i}
\tilde D_\rho \tilde X_i+3\tilde\lambda^2 \tilde C_{\mu ij}
\tilde D_\mu \tilde X_i \tilde D_\nu \tilde X_j+\tilde\lambda^3
\tilde C_{ijk}\tilde D_\mu \tilde X_i \tilde D_\nu \tilde X_j
\tilde D_\rho \tilde X_k\Big]\nonumber\\
&+\frac{\mu_2}{3!} \epsilon^{\mu\nu\rho}\Big[-\frac{i}{2}\tilde\lambda
\tilde C_{\mu\nu\rho ij8}[\tilde X_i,\tilde X_j]-\frac {3i}{2}
\tilde\lambda^2\tilde C_{\mu\nu ijk}\langle\hspace{-0.7mm}\langle
[\tilde X_i,\tilde X_j] \tilde D_\rho \tilde X_k\rangle
\hspace{-0.7mm}\rangle\nonumber\\
&-\frac {3i}{2}\tilde\lambda^3\tilde C_{\mu ijkl}\langle
\hspace{-0.7mm}\langle [\tilde X_i,\tilde X_j] \tilde D_\nu \tilde X_k
\tilde D_\rho \tilde X_l\rangle\hspace{-0.7mm}\rangle
-\frac {i}{2}\tilde\lambda^4\tilde C_{ijklm}\langle
\hspace{-0.7mm}\langle [\tilde X_i,\tilde X_j] \tilde D_\mu \tilde X_k
\tilde D_\nu \tilde X_l\tilde D_\rho \tilde X_m\rangle\hspace{-0.7mm}
\rangle\Big]\bigg\}\label{XCSM}.
\end{align}
The second and third terms in (\ref{XCSM}) are unified to form the kinetic
term of U(2) gauge field, as the gauge invariant combination on the
world-volume of D-brane,
\begin{align}
-\frac{1}{2g^2_{{\rm YM}}} {\rm Tr}\left[
\Big(\tilde F_{\mu\nu}+\frac 1{\tilde\lambda}
P[\tilde B_{\mu\nu}]\Big)\Big(\tilde F^{\mu\nu}+\frac 1{\tilde\lambda}
P[\tilde B^{\mu\nu}]\Big)\right]
\end{align}
up to the quadratic term in $P[\check B_{\mu\nu}]$ which belongs
to nonlinear terms in $C^{(3)}$.
Here we notice that in addition to the natural couplings
of the D2-brane to the three-form field and the dual  five-form
field in type IIA superstring theory, the WZ-type action also
produces the coupling between $\tilde{F}_{\mu\nu}$ and $\tilde{B}_{\mu\nu}$
in the linearized nonabelian DBI action for D2-brane. Unfortunately,
however, the dimensional reduction of the WZ-type action (\ref{liC})
does not produce $\tilde C^{(3)}\wedge\tilde F$
and $\tilde C^{(5)}\wedge\tilde F$-terms which appear in ten dimensional
WZ-type action (\ref{MCSA}). We need more investigations in this direction.

Our derivation of the result in (\ref{XCSM}) is based
on the elegant Higgs rule of \cite{Ezhuthachan:2009sr}.
Here we would like to comment on a mild problem in applying these rules.
We know that the transverse scalars $X_i$ are bi-fundamentals 
of SU(2)$\times$ SU(2).
Therefore, to obtain the U(2) adjoint scalar, the trace
and the traceless part of these scalars should be combined as
$\tilde X_i = \check x_i + {\bf x}_i$. A similar rewriting should also
be made for the form fields. When we are dealing with the BLG theory without
WZ-type coupling, the Higgs rule of \cite{Ezhuthachan:2009sr}
are exactly the net effect of this splitting and  recombination of the trace
and traceless part of the fields. However, in the presence of
the WZ-type coupling containing more than two covariant derivatives,
the splitting and recombination of the trace and traceless parts reexpress 
most of the terms of $S_C$ in terms of the U(2) adjoint
fields except a few terms which lead to some mismatch. 
To demonstrate this observation we
consider the $C_{ijk}$-term, 
\begin{align} \label{mmat}
&\frac{1}{2} \epsilon^{\mu\nu\rho}{\rm Tr}\left[ C_{ijk}
(\tilde D_\mu X_i)^\dagger
\tilde D_\nu X_j (\tilde D_\rho X_k)^\dagger + C^\dagger_{ijk} \tilde 
D_\mu X_i
(\tilde D_\nu X_j)^\dagger \tilde D_\rho X_k\right]
\nonumber \\
&=\epsilon^{\mu\nu\rho}{\rm Tr}\left(\tilde C_{ijk} \tilde D_\mu\tilde X_i
\tilde D_\nu\tilde X_j\tilde D_\rho\tilde X_k
-2 \tilde C_{ijk} \tilde D_\mu \check x_i\tilde D_\nu \check x_j
\tilde D_\rho x_k - 2 \tilde C_{ijk} \tilde D_\mu \check x_i\tilde D_\nu  x_j
\tilde D_\rho x_k\right),
\end{align}
where the covariant derivative is given in (\ref{tcov}). 
We have also made the following splitting and recombination
of the trace and traceless part of the three-form field
\begin{align}
&C_{ijk} = \frac{1}{2} \tilde c_{ijk} {\bf 1} + i c_{ijk}^\alpha
\frac{\sigma^\alpha}{2},
\quad C_{ijk}^\dagger = \frac{1}{2} \tilde c_{ijk} {\bf 1} - i c_{ijk}^\alpha
\frac{\sigma^\alpha}{2},
\quad \tilde C_{ijk} = \frac{1}{2} \tilde c_{ijk} {\bf 1} + c_{ijk}^\alpha
\frac{\sigma^\alpha}{2}.
\end{align}
Note that the the last two terms in (\ref{mmat})
cannot entirely be expressed in terms of the U(2) adjoint fields.
This mismatch is  generated from the cross
terms between the trace and traceless sectors.
It is quite straightforward to show that, in the
U(1)$\times$U(1) ABJM theory, there is no such mismatch.
We will leave verification of the absence of such mismatch
in ABJM theory with arbitrary gauge group for the future work~\cite{KKNT}.

\setcounter{equation}{0}
\section{Quartic Mass-Deformation Term from a $C_{(6)}$ Term}\label{sec4}

Let us recall the bosonic part of supersymmetry-preserving mass-deformation
terms in the BLG theory~\cite{Gomis:2008cv,Hosomichi:2008qk}
in order to compare these
with the WZ-type action in \eqref{MCSX},
\begin{align}
S_m=& \int d^3x\,
m^2 \, {\rm Tr}\Big(X_IX_I^\dagger+X_I^\dagger X_I\Big)+
\frac{4\pi m}{k} \int d^3x\, {\rm Tr}
\Big[X_3 (X_4)^\dagger X_5(X_6)^\dagger -X_5 (X_4)^\dagger X_3(X_6)^\dagger
\nonumber\\
&\hspace{75mm}+X_7 (X_8)^\dagger X_9(X_{10})^\dagger
-X_9 (X_8)^\dagger X_7(X_{10})^\dagger\Big],
\label{massdef}
\end{align}
where $m$ is the mass parameter.
According to~\cite{Bena:2000zb}, this mass term comes from the
background four-form flux which is (anti-)self-dual in
eight-dimensional transverse space.
We examine specifically how \eqref{massdef} can appear
from WZ-type coupling \eqref{liC}.
Due to the (anti-)self-dual property of the flux,
we should consider the contribution
from both four-form $F_{(4)}$ and dual seven-form $F_{(7)}$.
Let us first take into account the contribution from $F_{(7)}$ by turning on
only specific components of $F_{(7)}$ as
\begin{align}
F_{\mu\nu\rho IJKL}=\frac{\beta m}{\lambda\mu'_2}
\epsilon_{\mu\nu\rho}T_{IJKL},
\label{fcon}
\end{align}
where $T_{IJKL}$ is (anti-)self-dual in eight-dimensional transverse space
\begin{align}
T^{IJKL}=\pm\frac{1}{4!}\epsilon^{IJKLI'J'K'L'}T_{I'J'K'L'}.
\label{8sd}
\end{align} 

The corresponding WZ-type action of our consideration \eqref{MCSX} is
\begin{align}
S_{C}^{(6)} =\int dx^\mu\wedge dx^\nu\wedge dx^\rho~
\frac{\mu'_2}{2}{\rm Tr}\Big[C_{\mu\nu\rho IJK}(X_I)^\dagger X_J
(X_K)^\dagger+(\mbox{c.c.)}\Big]+\cdots\, .
\label{massterm}
\end{align}
When the six-form $C_{\mu\nu\rho IJK}$
is the potential of constant seven-form field strength $F_{\mu\nu\rho IJKL}$, 
we obtain it explicitly, 
\begin{align}
C_{\mu\nu\rho IJK}=\lambda F_{\mu\nu\rho IJKL}X_L
=\frac{\beta m}{\mu'_2}~
\epsilon_{\mu\nu\rho}T_{IJKL}X_L .
\label{cco}
\end{align}
Substituting the six-form field configuration \eqref{cco} with \eqref{8sd}
into the WZ-type action \eqref{massterm}, we have
\begin{align}
S_{C}^{(6)} & =\frac{4\pi m}{3k}\int dx^\mu\wedge dx^\nu\wedge
dx^\rho\epsilon_{\mu\nu\rho}
\frac 12{\rm
Tr}\Big[T_{IJKL}X_L(X_I)^\dagger X_J
(X_K)^\dagger +(\mbox{c.c.})\Big]+\cdots\nonumber\\
&=-\frac{4\pi m}{k}\int d^3x\,
{\rm Tr}\Big[T_{IJKL}X_L(X_I)^\dagger
X_J (X_K)^\dagger
+(\mbox{c.c.})\Big]+\cdots,
\label{massterm2}
\end{align}
which exactly coincides with the quartic mass-deformation term in
\eqref{massdef} as far as the four-form tensor $T_{IJKL}$ satisfies
\begin{align}
T_{1234}=T_{5678}=1,
\quad \mbox{other independent components}=0.
\label{spmd}
\end{align}
This configuration keeps the maximal supersymmetry ${\cal N}=8$, and
turning on other components of $T_{IJKL}$ in addition to \eqref{spmd}
leads to less supersymmetry. The relation between nonzero components of
$T_{IJKL}$ and the number of supersymmetry has been studied in the context of
field theory~\cite{Hosomichi:2008qk,Ahn:2008gda,Nishino:2008zzb}
and in the dual AdS side~\cite{Bena:2000zb}.

The contribution from the four-form tensor is also calculated
in a similar way. From \eqref{fcon} the configuration of $F_{(4)}$ is
\begin{align}
F_{IJKL}\sim \frac{\beta m}{\lambda\mu'_{2}}T_{IJKL}
=\frac{m}{\lambda^{2}\mu_{2}}T_{IJKL}.
\end{align}
Then the corresponding WZ-type action is
\begin{align}
S_{C}^{(3)} &\sim \mu_{2}\int dx^\mu\wedge dx^\nu\wedge
dx^\rho \,
{\rm Tr}\biggl[\frac{m}{\lambda^{2}\mu_{2}}T_{IJKL}
\lambda^{4}X_L(D_{\mu}X_I)^\dagger D_{\nu}X_J(D_{\rho}X_K)^\dagger
+(\mbox{c.c.})\biggr]+\cdots
\notag\\
&=\lambda^{2}m\int dx^\mu\wedge dx^\nu\wedge
dx^\rho \,
{\rm Tr}\Bigl[T_{IJKL}
X_L(D_{\mu}X_I)^\dagger D_{\nu}X_J(D_{\rho}X_K)^\dagger
+(\mbox{c.c.})\Bigr]+\cdots.
\label{massterm3}
\end{align}
Since $\lambda^{2}\sim l_{{\rm P}}^{3}$, this term does not contribute
under the limit $l_{{\rm P}}\to 0$.

We confirm the identification of a cubic WZ-type term (\ref{massterm}) with a
specific form of constant (anti-)self-dual four-form flux (\ref{fcon})
in Minkowski spacetime signature. If we take a Euclideanization to the flux
(\ref{fcon}), then an overall imaginary number $i$ appears in the left-hand
side. It implies that $\mu$ in (\ref{fcon}) may not be a mass parameter
but a chemical potential.

For the quadratic mass term which can be interpreted as
the quadratic coupling of form fields between M2's,
we do not have natural argument to fix it.
In the case of string theory, this coupling is obtained
from the world-sheet disk amplitude with insertion of
the two R-R vertex operators. Though we basically have ambiguity for the
position of two insertions, we can avoid this ambiguity by introducing
appropriate auxiliary fields~\cite{Atick:1987gy,Dine:1987gj}
and compute the coupling at least for some particular
cases~\cite{Billo:2004zq,Ito:2006ig}.

\setcounter{equation}{0}
\section{Conclusion and Discussion}\label{sec5}

Once the world-volume action of $N$ stacked M2-branes is determined,
it is interesting to understand how the M2-branes couple to
the bulk fields. In this paper, we constructed the WZ-type action which
describes the coupling of the M2-branes to antisymmetric three-
and six-form fields in M-theory.
We consider the BLG theory for two M2-branes and write down a WZ-type action
linear to antisymmetric three- and six-form fields
in analogy with the corresponding action in type IIA string theory.
When it reduces to ten dimensions through a circle compactification,
our action reproduces the expected ten-dimensional coupling of R-R and
NS-NS form fields to D2-branes in type IIB string theory.

In addition to our main goal of obtaining the WZ-type coupling of M2-branes,
we show that a particular cubic WZ-type term can be identified with the
quartic scalar interaction in the supersymmetry-preserving mass deformation
of the BLG theory. We made this identification in a flat world-volume
and transverse space by making an assumption that the seven-form field
strength is constant and is proportional to the mass parameter of mass
deformation term.

A few discussions are in order.
Though the ten-dimensional WZ-type action \eqref{MCSA1}
is restricted to the terms linear in $\tilde C_{(n)}$,
it contains $\tilde C_{(n)}\wedge e^{\tilde B}$ where $\tilde B$
is NS-NS two-form field.
Since the NS-NS two-form field in string theory comes from a part of
$C_{(3)}$ in M-theory, inclusion of quadratic or higher order terms
in $C_{(3)}$ in addition to the WZ-type action \eqref{MCSX} seems natural.
For instance the quadratic term in $C_{(3)}$ is
\begin{align}
{S_{C^2}} =&\frac{\mu_2''}2\int_{2+1}{\rm Tr}\Big(P[
\langle{\rm i}_X{\rm i}_X{\rm i}_X\rangle
C_{(3)}
\wedge C_{(3)}]\Big)
\nonumber\\
=&5\mu_2''\int\frac 1{3!} d^3x \epsilon^{\mu\nu\rho}~
{\rm Tr}\bigg[
\frac{1}{2}\hat C_{[\mu\nu\rho} \langle\hspace{-0.7mm}\langle
C^\dagger_{IJK]} X_{IJK}\rangle\hspace{-0.7mm}\rangle +
\frac{1}{2} C_{[\mu\nu\rho} \langle\hspace{-0.7mm}\langle
C_{IJK]} X_{IJK}^\dagger\rangle\hspace{-0.7mm}\rangle
\nonumber\\
&~~~~~~~~~~~~~~~~~~\hspace{10mm}
+3\lambda C_{[\mu\nu L}\langle\hspace{-0.7mm}\langle C^\dagger_{IJK]}
 X_{IJK}(D_\rho X_L)^\dagger \rangle\hspace{-0.7mm}\rangle
\nonumber\\
&~~~~~~~~~~~~~~~~~~\hspace{10mm}
+\frac{3}{2}\lambda^2\hat C_{[\mu LM}
\langle\hspace{-0.7mm}\langle C^\dagger_{IJK]} X_{IJK} (D_\nu X_L)^\dagger
D_\rho X_M \rangle\hspace{-0.7mm}\rangle
\nonumber\\
&~~~~~~~~~~~~~~~~~~\hspace{10mm}
+\frac{3}{2}\lambda^2 C_{[\mu LM} \langle\hspace{-0.7mm}\langle
C_{IJK]} X^\dagger_{IJK} D_\nu X_L (D_\rho X_M)^\dagger
\rangle\hspace{-0.7mm}\rangle
\nonumber\\
&~~~~~~~~~~~~~~~~~~\hspace{10mm}
+\lambda^3 C_{[LMN} \langle\hspace{-0.7mm}\langle
C^\dagger_{IJK]}X_{IJK}(D_\mu X_L)^\dagger D_\nu X_M
(D_\rho X_N)^\dagger \rangle\hspace{-0.7mm}\rangle
+(\mbox{c.c.})\bigg],
\label{CC}
\end{align}
where $\mu_2''= \beta' \lambda\mu_2$.
Like $\beta$, the value of $\beta'$ is also determined by comparing
this term  with an appropriate
term in the corresponding action of
type IIA superstring theory, after a circle compactification. 
Specifically, when $\beta'=\beta=4\pi/3k$,
this term exactly coincides with the $C_{(3)}\wedge B$ term in (\ref{MCSA1}).

In this paper we constructed the bosonic sector of WZ-type action coupled to
the world-volume fields of M2-branes. If we supersymmetrize what we obtained,
then we may reach the supersymmetric WZ-type action in M-theory.
Though we fixed the coefficient of the WZ-type action \eqref{MCSX} by
comparing with the terms of the ten-dimensional R-R coupling
action \eqref{MCSA1} through the compactification of the eighth transverse
direction, this indirect fixation procedure can be reconfirmed by constructing
the supersymmetric WZ-type action coupled to M2-branes.
Then, this understanding will also help the extension to general case of
arbitrary number of stacked multiple M2-branes of which the world-volume
theory is described by the ${\cal N}=6$ superconformal Chern-Simons gauge
theory with U($N)\times$U($N$) gauge symmetry.
We will report the construction of WZ-type coupling in the context of
ABJM theory for arbitrary number of M2-branes in the subsequent
work~\cite{KKNT}.

In the BLG and ABJM theories, M2-branes and ${\bar{\rm M}}$2-branes are not
distinguished as the case of DBI type world-volume action of D-branes.
The D- and ${\rm \bar D}$-branes carry opposite sign R-R charges and
are distinguished by the R-R coupling (\ref{MCSA1}) in type II string
theories~\cite{Myers:1999ps}. Similarly the M2- and ${\rm \bar M2}$-branes
are also distinguishable by the analogue of WZ-type in the M-theory
(\ref{MCSX}). This will also let the construction of world-volume action of
M2${\rm \bar M2}$ pair without supersymmetry tractable.

\section*{Acknowledgements}
The authors would like to appreciate the informative discussions with
Min-Young Choi and Akira Ishida.
This work was supported by the Korea Research Foundation Grant funded by
the Korean Government with grant number KRF-2008-313-C00170 (Y.K.), 
2009-0073775 (O.K.), and 2009-0077423 (D.T.).
This work was also supported by Astrophysical Research
Center for the Structure and Evolution of the Cosmos (ARCSEC)).

\end{document}